\documentclass[twocolumn]{aastex63}
\usepackage{amsmath}

\shorttitle{Relativistic Electron-Positron Wind}
\shortauthors{Li \& Dai}

\begin{document}

\title{GRB 170817A Afterglow from a Relativistic Electron-Positron Pair Wind Observed Off-axis}

\author[0000-0002-8391-5980]{Long Li}
\affiliation{School of Astronomy and Space Science, Nanjing University, Nanjing 210023, China}
\affiliation{Key Laboratory of Modern Astronomy and Astrophysics (Nanjing University), Ministry of Education, China}

\author[0000-0002-7835-8585]{Zi-Gao Dai}
\affiliation{Department of Astronomy, University of Science and Technology of China, Hefei 230026, China; daizg@ustc.edu.cn}
\affiliation{School of Astronomy and Space Science, Nanjing University, Nanjing 210023, China}

\begin{abstract}

A relativistic electron-positron ($e^{+}e^{-}$) pair wind from a rapidly rotating, strongly magnetized neutron star (NS) would interact with a gamma-ray burst (GRB) external shock and reshapes afterglow emission signatures. Assuming that the merger remnant of GW170817 is a long-lived NS, we show that a relativistic $e^{+}e^{-}$ pair wind model with a simple top-hat jet viewed off-axis can reproduce multi-wavelength afterglow lightcurves and superluminal motion of GRB 170817A. The Markov chain Monte Carlo (MCMC) method is adopted to obtain the best-fitting parameters, which give the jet half-opening angle $\theta_{j}\approx0.11$ rad, and the viewing angle $\theta_{v}\approx0.23$ rad. The best-fitting value of $\theta_{v}$ is close to the lower limit of the prior which is chosen based on the gravitational-wave and electromagnetic observations. In addition, we also derive the initial Lorentz factor $\Gamma_{0}\approx47$ and the isotropic kinetic energy $E_{\rm K,iso}\approx2\times10^{52}\rm\ erg$. A consistence between the corrected on-axis values for GRB 170817A and typical values observed for short GRBs indicates that our model can also reproduce the prompt emission of GRB 170817A. An NS with a magnetic field strength $B_{p}\approx1.6\times10^{13}\rm\ G$ is obtained in our fitting, indicating that a relatively low thermalization efficiency $\eta\lesssim10^{-3}$ is needed to satisfy observational constraints on the kilonova. Furthermore, our model is able to reproduce a late-time shallow decay in the X-ray lightcurve and predicts that the X-ray and radio flux will continue to decline in the coming years.

\end{abstract}

\keywords{Gamma-ray bursts(629); Gravitational waves(678); Hydrodynamics(1963); Non-thermal radiation sources(1119); Neutron stars(1108)}

\section{Introduction} \label{sec:intro}

On 2017 August 17 at 12:41:04 UTC, the Advanced Laser Interferometer Gravitational-Wave Observatory and Advanced Virgo Interferometer gravitational-wave detectors detected the first gravitational-wave (GW) event GW170817 from a binary neutron star (BNS) merger \citep{Abbott2017a}.
About 1.7 seconds after the coalescence, \emph{Fermi} Gamma-ray Burst Monitor and \emph{International Gamma-Ray Astrophysics Laboratory} were triggered by an low-luminosity short gamma-ray burst (sGRB) GRB 170817A, independently \citep{Abbott2017c,Goldstein2017,Savchenko2017}. The detection of GRB 170817A confirmed that at least some sGRBs are associated with BNS merger events.
Roughly 11 hours later, an optical counterpart AT2017gfo was discovered to be around the galaxy NGC 4993 \citep{Coulter2017}. Counterparts at ultraviolet and infrared band were also detected within one day \citep{Soares-Santos2017}. These observations support the hypothesis that AT 2017gfo is a kilonova powered by the radioactive decay of rapid neutron-capture process nuclei synthesized within the ejecta \citep{Kasen2017,Pian2017}.
About 9 days later, Chandra X-ray Observatory detected X-ray emission at the position of AT2017gfo \citep{Troja2017}. A longer time later, a radio band counterpart was detected \citep{Hallinan2017}. The X-ray and radio emission can be characterized by a non-thermal power-law spectrum, which is consistent with the GRB afterglow from a relativistic shock.
The gravitational-wave signal combined with following electromagnetic counterparts helps to understand the process of a BNS merger and possible products, showing a breakthrough for multi-messenger astronomy.

Continued follow-up observations reveal that the luminosity of a multi-wavelength afterglow was gradually increasing, peaked around 160 days after the merger, and then started to rapidly decreasing \citep{Margutti2018,Mooley2018a,Ruan2018,Makhathini2020,Troja2020}.
The multi-wavelength behavior from radio to X-rays is different from a typical GRB afterglow. Considering the low luminosity of the prompt emission, the explanations for the afterglow luminosity evolution usually fall into two types. One explanation is invoking a structured jet which have an angular distribution in energy and Lorentz factor. The GRB 170817A afterglow is well consistent with the emission from structured jet viewed off-axis \citep{Abbott2017b,Troja2017,Lazzati2018,Lyman2018,Margutti2018,Mooley2018b,Troja2018,Hajela2019,Lamb2019,Troja2019}.
The second explanation is based on a classical top-hat jet model with considering additional continuous energy injection into a jet (e.g. \citealt{Geng2018,Li2018,Lamb2020}).

On the other hand, the central remnant of GW170817 which depends on the neutron star (NS) equation of state (EoS) remains up in the air. Depending on whether the NS EoS is stiff enough, the remnant could be a black hole, a temporal hypermassive NS, or a long-lived massive NS. Although there are theories against a long-lived massive NS as the merger remnant (e.g. \citealt{Granot2017,Margalit2017,Ciolfi2020}), there is no direct evidence that rules out the possibility that the post-merger remnant is a stable NS. A relativistic jet could also be launched through $\nu\bar{\nu}$ annihilation in a neutrino-dominated accretion flow (NDAF) mechanism from such a stable NS with a hyper-accreting accretion disk \citep{Zhang2008,Zhang2009,Zhang2010}. Besides, the existence of an NS remnant can reduces the requirement on the ejecta mass in the kilonova model due to an additional energy injection from the NS \citep{Li2018,Metzger2018,Yu2018}. If the remnant is a long-lived massive NS, it is suggested that a Poynting flux-dominated outflow would flow out from the NS. This outflow can be accelerated due to magnetic dissipation (e.g. reconnection), and eventually dominated by the energy flux of ultra-relativistic wind consists of electron-positron ($e^+e^-$) pairs \citep{Coroniti1990,Michel1994,Kirk2003}. \cite{Dai2004} realized that the continuous ultra-relativistic $e^+e^-$ pair wind would interact with the GRB external shock and reshape the afterglow emission signatures. \cite{Yu2007} used this model to explain the shallow decay phase of the early GRB X-ray afterglows. \cite{Geng2018} modeled the first 150 days of GRB 170817A afterglow lightcurves in this scenario. However, some new observational facts have been updated since then. Soon afterwards, \cite{Mooley2018c} reported the radio observations of GRB 170817A afterglow using a Very Long Baseline Interferometry (VLBI), and found that the radio source shows superluminal apparent motion between 75 days and 230 days after the merger event. A mean apparent velocity of the radio source along the plane of the sky $v_{\rm app} = (4.1\pm0.5)c$ was measured. Recently, Chandra X-ray Observatory continuous detected X-ray emission from the location of GRB 170817A \citep{Hajela2020a,Hajela2020b,Hajela2021a,Hajela2021b}, which extended the afterglow data to about 3.3 years after the merger. The late-time derived unabsorbed X-ray flux is higher than what is expected from the structured jet model \citep{Makhathini2020,Troja2020,Hajela2020b,Hajela2021a,Hajela2021b}. However, the late-time radio observations did not show this excess \citep{Balasubramanian2021,Hajela2021b}. Therefore, it is necessary to revisit an off-axis afterglow from an $e^+e^-$ pair wind scenario supposing that the remnant of GW170817 is a long-lived massive NS.

This paper is organized as follows. In subsection \ref{sec:model} we give the details of our numerical afterglow model from an $e^+e^-$ pair wind scenario. In subsection \ref{sec:170817} we describe the methods of fitting the data from GRB 170817A with our model. Our fitting data includes the thousand-day multi-wavelength afterglow and the VLBI proper motion. In Section \ref{sec:res} we describe our results. Our discussion and conclusions can be found in Section \ref{sec:discon}. A concordance cosmology with parameters $H = 69.6 \rm\ km\ s^{-1}\ Mpc^{-1}$, $\Omega_{M} = 0.286$, and $\Omega_\Lambda = 0.714$ is adopted in this paper.

\section{Method} \label{sec:meth}

\subsection{The relativistic $e^{+}e^{-}$ pair wind model} \label{sec:model}

We assume that the central remnant of GW170817 is a long-lived NS. A Poynting-flux-dominated outflow is launched from the NS, whose wind luminosity is dominated by the magnetic dipole spin-down luminosity, i.e. \citep{Shapiro1983}
\begin{eqnarray}
L_{w} = \frac{B_{p}^{2} R_{s}^{6} \Omega^{4}}{6 c^{3}} & \simeq & 9.64 \times 10^{44} B_{p,13}^{2} P_{0,-3}^{-4} R_{s,6}^{6}\nonumber\\ & & \times \left(1+\frac{t_{\rm obs}}{\tau_{\rm sd}}\right)^{-2} \rm\ erg\ s^{-1},
\end{eqnarray}
where $B_{p}=10^{13}B_{p,13}\,{\rm G}$, $P_{0}=10^{-3}P_{0,-3}\,{\rm s}$, and $R_{s}=10^{6}R_{s,6}\,{\rm cm}$ are the polar surface magnetic field strength, radius, and initial spin period of NS, respectively. $t_{\rm obs}$ is the time measured in the observer frame. $\tau_{\rm sd} = 2.05 \times 10^{7} I_{45} B_{\rm p,13}^{-2} P_{\rm 0,-3}^{2} R_{\rm s, 6}^{-6} \rm\,s$ is the characteristic spindown time scale, where $I=10^{45}I_{45}\,{\rm g\ cm^{2}}$ is the moment of inertia of the NS. For simplicity, $P_{0,-3}$, $R_{s,6}$, and $I_{45}$ are taken to be unity throughout this work.

As it propagates outwards, the Poynting-flux dominated outflow can be dissipated and accelerated by magnetic reconnection, which is eventually dominated by an ultra-relativistic wind consisting of $e^+e^-$ pairs with bulk Lorentz factor $\Gamma_{w} \sim 10^4-10^7$ \citep{Atoyan1999}. We here adopt $\Gamma_{w}=10^4$ as a fiducial value in our calculations. Based on the luminosity $L_{w}$ and the bulk Lorentz factor $\Gamma_{w}$ of the $e^+e^-$ pair wind, and the assumption that the $e^{+}e^{-}$ pair wind is isotropic, the comoving electron number density can be estimated as
\begin{equation}
n_{w}^\prime \simeq \frac{L_{w}}{4 \pi R^{2} \Gamma_{w}^{2} m_{e} c^{3}},
\end{equation}
where $m_{e}$ is the mass of electron, $c$ is the speed of light, and $R$ is the distance from the central engine.

\begin{figure}
\centering
\includegraphics[width=0.4\textwidth,angle=0]{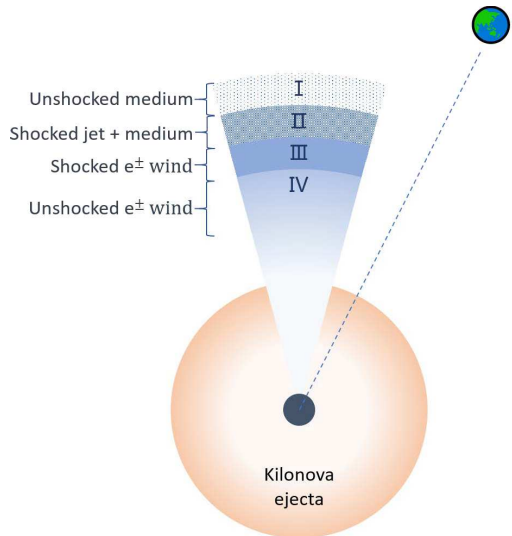}
\caption{A cartoon picture of the relativistic $e^+e^-$ pair wind model with a top-hat jet viewed off-axis, if the post-merger remnant is a rapidly spinning magnetized NS.
\label{fig:cart}}
\end{figure}

As shown in Figure \ref{fig:cart}, most of the wind energy is injected into the kilonova ejecta, only a small fraction (($1-\cos\theta_j$)/2 where $\theta_j$ is the half opening angle of jet) of the $e^+e^-$ pair wind propagates outside the ejecta in the jet direction. when the ultra-relativistic $e^+e^-$ pair wind interacts with its surrounding medium, a pair of shocks will develop: a forward shock (FS) that propagates into the medium, and a reverse shock (RS) that propagates into the wind. Before this interaction, a forward shock has formed when the GRB jet interacts with the medium. For simplicity, we here assume that the two forward shocks eventually merged into one forward shock. Thus, there are four regions separated by two shocks: (1) the unshocked medium, (2) the shocked jet and medium, (3) the shocked wind, and (4) the unshocked wind \citep{Dai2004}. Regions 2 and 3 are separated by the contact discontinuity.

We solve a set of differential equations which describe the dynamics of such an FS-RS system and consider both synchrotron radiation and inverse Compton (IC) radiation from shock-accelerated electrons. The details of calculating the multi-wavelength lightcurves can be seen in Appendix \ref{app:dyn}.

\subsection{Modeling the GRB 170817A afterglow} \label{sec:170817}

A numerical model is established to fit the GRB 170817A afterglow data including the multi-wavelength afterglow and the VLBI proper motion. The multi-wavelength afterglow data are taken from the following literature: \cite{Makhathini2020} collected and reprocessed the available radio, optical and X-ray data spanning from 0.5 days to 1231 days after merger, \cite{Balasubramanian2021} updated the radio observations to about 3.5 years after merger, and \cite{Hajela2020b,Hajela2021a} updated the 0.3-10 keV X-ray data to 3.3 years after merger. Our fitting data include following specific bands: frequencies at 3 GHz and 6 GHz from Karl G. Jansky Very Large Array (VLA), wavelength at 600nm from Hubble Space Telescope (HST) F606W, energy at 1 keV from Chandra X-ray Observatory and XMM-Newton Observatory, and energy at 0.3-10 keV from Chandra X-ray Observatory, which can be regarded as representative of the multi-wavelength afterglow, as shown in Figure \ref{fig:multi}. In order to fit the VLBI proper motion, we develop a method used to calculate the proper motion of the flux centroid, which can be seen in Appendix \ref{app:prop}. As part of fitting data, we construct a data point based on the mean apparent velocity of the source, as shown in Figure \ref{fig:beta}.

We develop a Fortran-based numerical model and implement the Markov Chain Monte Carlo (MCMC) techniques by using the \texttt{emcee} Python package \citep{emcee}. \texttt{F2PY} is employed to provide a connection between Python and Fortran languages. We perform the MCMC with 22 walkers for running at least 100,000 steps, until the step is longer than 50 times the integrated autocorrelation time $\tau_{\rm ac}$, i.e. $N_{\rm step}=\max(10^5, 50\tau_{\rm ac})$ to make sure that the fitting is sufficiently converged \citep{emcee}. Once the MCMC is done, the best-fitting values and the $1\sigma$ uncertainties are computed as the 50th, 16th, and 84th percentiles of the posterior samples.

The free parameters in the fitting include: the half opening angle of jet ($\theta_{j}$), the viewing angle ($\theta_{v}$), the NS surface magnetic field strength at the polar cap region ($B_{p}$), the isotropic kinetic energy of the jet after prompt emission ($E_{\rm K,iso}$), the initial Lorentz factor of the jet after prompt emission ($\Gamma_{0}$), the number density of medium ($n_{1}$), the fraction of the shock internal energy that is partitioned to magnetic fields in regions 2 ($\epsilon_{B,2}$) and 3 ($\epsilon_{B,3}$), the fraction of the shock internal energy that is partitioned to electrons in region 2 ($\epsilon_{e,2}$), the electron energy spectral index in regions 2 ($p_{2}$) and 3 ($p_{3}$). The $\epsilon_{e,3}$ is not regarded as a free parameter because the implicit condition $\epsilon_{e,3}+\epsilon_{B,3}=1$ in region 3 \citep{Dai2004,Yu2007,Geng2016,Geng2018}.

Uniform priors on $\theta_{j}$, $\theta_{v}$, $\log B_{p}$, $\log E_{\rm K,iso}$, $\log n_{1}$, $\log\epsilon_{B,2}$, $\log\epsilon_{B,3}$, $\log\epsilon_{e,2}$, $p_{2}$, and $p_{3}$ are adopted in this paper. In order to explore a parameter space as large as possible, we set a wide enough range for the priors, except for $\theta_{v}$. \cite{Abbott2017a} derived $\theta_{v}\leq56^{\circ}$ at 90\% credible intervals with a low spin prior by using the distance measured independently by GW.
\cite{Finstad2018} obtained $\theta_{v}$ with the distance measurement from \cite{Cantiello2018} and the GW data from \cite{Abbott2017a}. They derived the 90\% confidence region on the viewing angle is $\theta_{v}=32_{-13}^{+10}\pm1.7$ degrees, and derived a conservative lower limit on the viewing angle $\theta_{v}\geq13^{\circ}$. \cite{Abbott2019} derived the binary inclination angle $\theta_{JN}=151_{-11}^{+15}$ degrees\footnote{Supposing the jet is aligned with the the angular momentum, one can easily connect $\theta_{v}$ to the binary inclination angle $\theta_{JN}$ according to $\theta_{v}\equiv\min(\theta_{JN}, 180^{\circ}-\theta_{JN})$.} with a low spin prior by jointing GW and electromagnetic (EM) observations. Conservatively, we limit the prior range for $\theta_{v}$ to $0.23\leq\theta_{v}\leq0.7$ (labeled Prior A). As a comparison, we also show the fitting results from priors with $\theta_v\in[0,0.7]$ (labeled Prior B). The free parameters and priors in our model can be seen in Table \ref{tab:pars}.

\section{Results} \label{sec:res}

\begin{figure}
\centering
\includegraphics[width=0.5\textwidth,angle=0]{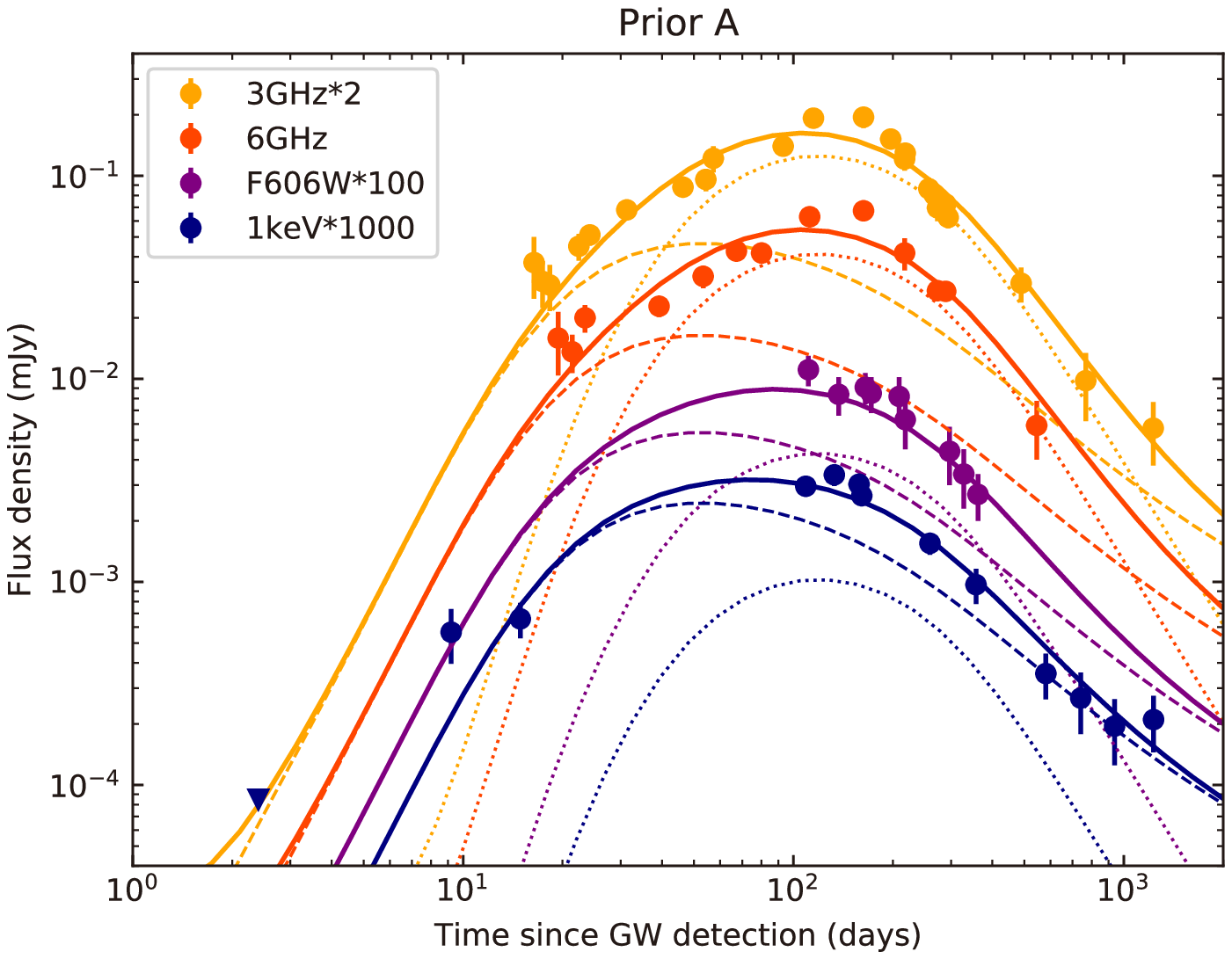}
\includegraphics[width=0.5\textwidth,angle=0]{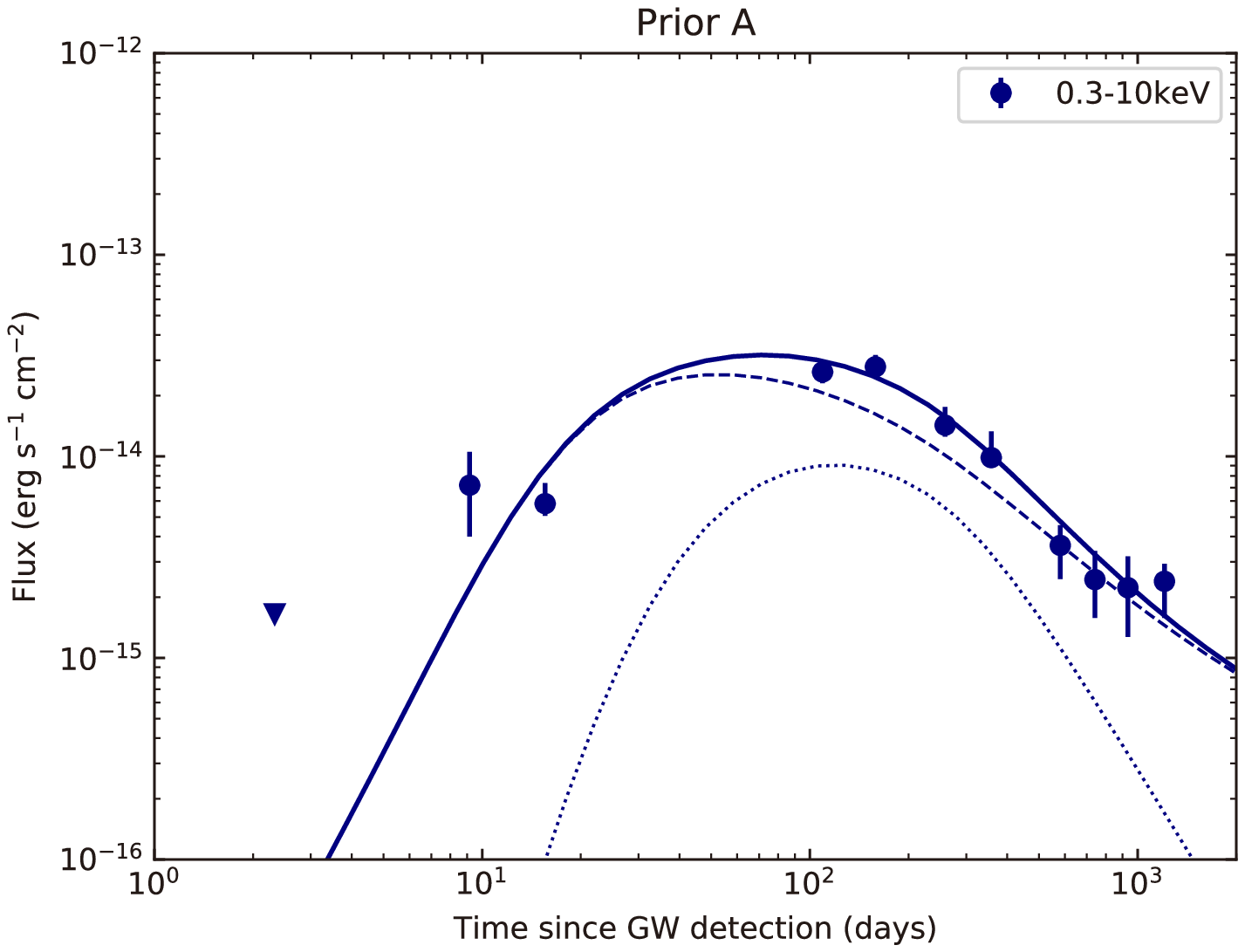}
\caption{Relativistic $e^+e^-$ pair wind model fit to the multi-wavelength afterglow lightcurves of GRB 170817A. The dashed and dotted curves represent the emission from the FS and RS, respectively. The solid curves are the total emission lightcurves.
\label{fig:multi}}
\end{figure}
\begin{figure*}
\centering
\includegraphics[width=1.0\textwidth,angle=0]{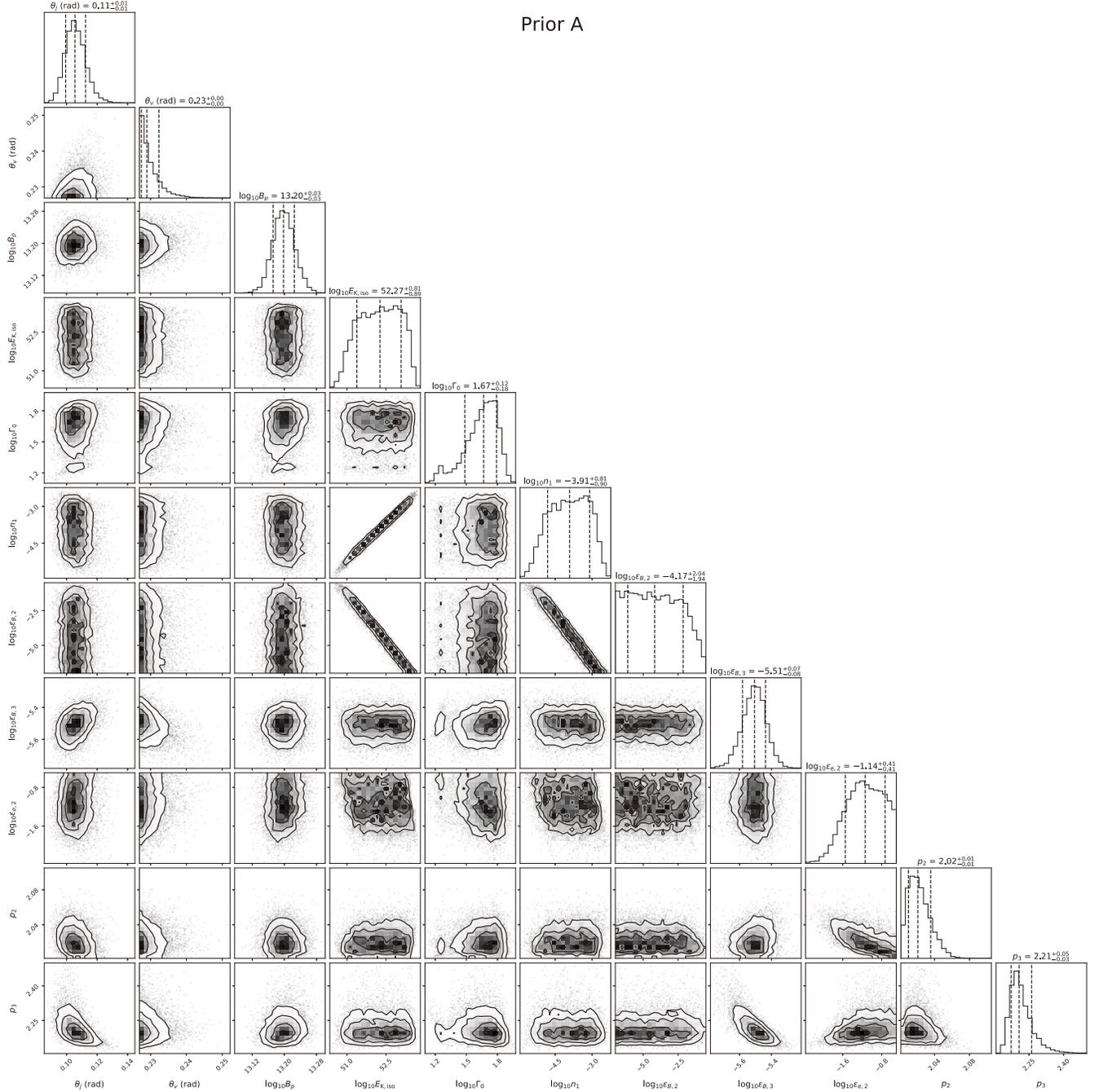}
\caption{A corner plot showing the results of our MCMC parameter estimation for the relativistic $e^+e^-$ pair wind model. Our best-fitting parameters and corresponding $1\sigma$ uncertainties are shown with the black dashed lines in the histograms on the diagonal.
\label{fig:corner}}
\end{figure*}

Figures \ref{fig:multi} and \ref{fig:beta} show the data of the afterglow, and the best-fitting results from Prior A. In Appendix \ref{app:PB}, we show the fitting results from Prior B. We find that for both priors the overall quality of the fitting is good, except for the early-time X-ray lightcurve which exhibit some deviation from the first data point. Table \ref{tab:pars} shows the $\chi_{\rm{dof}}^2\approx3$ and $\approx2$ for Prior A and B, indicating that a large parameter space for $\theta_v$ would improves the fitting quality. The corner plots shown in Figure \ref{fig:corner} indicate that the posterior distribution of $\theta_v$ close to the lower limit of its prior, which means a smaller $\theta_v$ would lead to a better fit. After performing a full prior of $\theta_v$ ranging from 0 to 0.7, the posterior distribution of $\theta_v$ exhibits a normal distribution peaking at about 0.14 rad ($\approx8$ deg), as shown in Figure \ref{fig:corner2}. The derived $\theta_{v}\approx8$ deg from Prior B is consistent with the viewing angle inferred from the detection of GW170817 ($\theta_v\leq55$ deg at 90\% confidence with a low-spin priors; \citealt{Abbott2017a}), but lower than the viewing angle inferred by combining GW and EM constraints (usually $\gtrsim10$ deg; \citealt{Finstad2018,Mandel2018,Abbott2019}). The smaller $\theta_v$ may pose a greater challenge to the estimation of the Hubble flow velocity for NGC 4993 or the constraint on Hubble constant $H_0$. Considering the reasonably good fit from Prior A, we conclude the results from Prior A are more reasonable than those from B. As shown in Appendix \ref{app:PB}, the results from Prior B are similar to Prior A, except for the derived smaller $\theta_j$ and $\theta_v$, and the difference in the apparent source size which we describe below. In the following we only discuss the results from Prior A.

In particular, our model can well reproduce the slowly rising phase of the early-time lightcurves. Instead of invoking a structured jet, our model suggests that the initial slowly rising lightcurves are a consequence of the peak time difference between the FS emission and RS emission. The FS emission reaches to a peak when the Lorentz factor of the jet $\Gamma\sim1/(\theta_v-\theta_j)$, while the peak time of the RS emission is roughly equal to the spin-down time scale of the NS \citep{Geng2016}. In our fitting, the peak times of FS and RS emission are about 50 days and 150 days respectively, and the superposition of these two components flatten the radio$-$X-ray lightcurves during this period, although emission of each component rises sharply. The late-time X-ray afterglow of GRB 170817A shows a clear excessive emission as compared with the estimated emission from the off-axis structured jet model \citep{Hajela2021b}. This excess could be explained by invoking additional emission component (e.g. kilonova afterglow; \citealt{Hajela2019,Troja2020,Hajela2021b}, accretion-powered emission; \citealt{Hajela2021b,Ishizaki2021}, or energy injection from the NS; \citealt{Troja2020,Hajela2021b}). Our model can reproduce a late-time shallower decay (but cannot reproduce a late-time rise; as shown in Figure \ref{fig:multi}) in the X-ray lightcurve without invoking an additional energy source. Moreover, our model can give a natural explanation of the late-time harder spectrum \citep{Hajela2021b} due to the change of composition and a relatively small $p$ from the FS (the transition of the blast wave dynamics from the relativistic phase to the sub-relativistic phase can also lead to a shallower decay in the late-time lightcurves without spectral evolution).

\begin{figure}
\centering
\includegraphics[width=0.5\textwidth,angle=0]{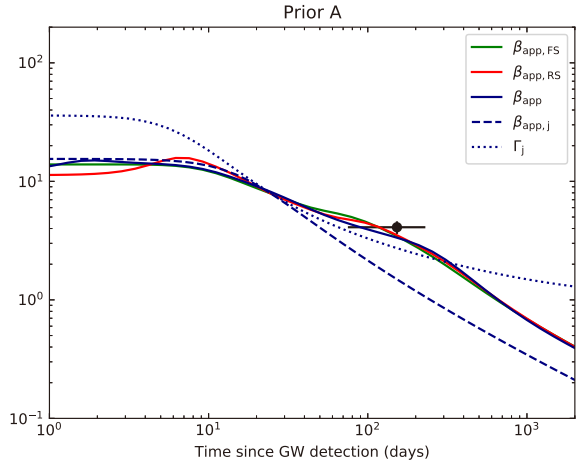}
\caption{Relativistic $e^+e^-$ pair wind model fit to the VLBI proper motion of GRB 170817A afterglow. The mean dimensionless apparent velocity of the radio afterglow source $\beta_{\rm app}=4.1\pm0.5$ between 75 days and 230 days after the merger from \cite{Mooley2018c} yields the data point used to fit. The green, red, and blue solid curves show the evolution of dimensionless apparent velocity of the flux centroid based on our best-fitting parameters when considering FS emission, RS emission and total emission, respectively. The blue dashed and dotted curves denote $\beta_{\rm app,j}$ and $\Gamma_{\rm j}$.
\label{fig:beta}}
\end{figure}

\begin{figure}
\centering
\includegraphics[width=0.5\textwidth,angle=0]{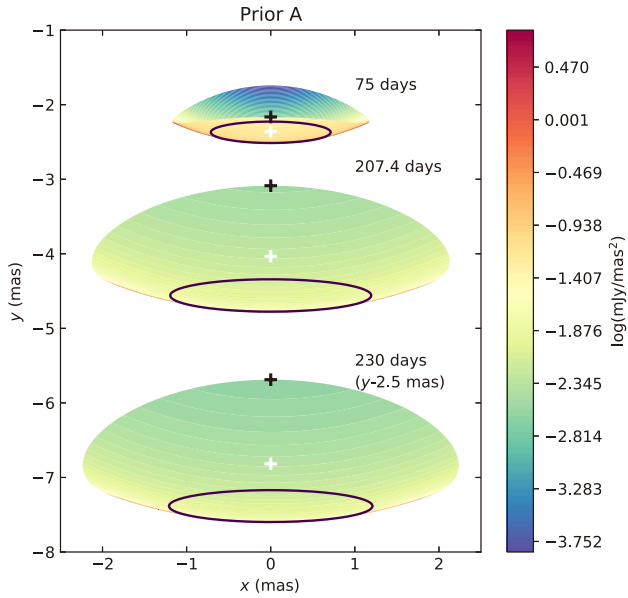}
\caption{The relative position of 4.5 GHz radio images at 75 days (upper), 207.4 days (middle), and 230 days (lower) seen by the observer. The white plus sign is the location of the flux centroid. The black plus sign is the location closest to the LOS. Purple ellipses are the best-fitting elliptical Gaussian showing the sizes (full width at half maximum) of the images.
\label{fig:image}}
\end{figure}

Figure \ref{fig:beta} shows the best-fitting results for apparent velocity. A consistency between the model and the data suggests that our model can also reproduce the superluminal apparent motion between 75 days and 230 days after the merger. We also plot the dimensionless apparent velocity $\beta_{\rm app,j}$ and Lorentz factor $\Gamma_{\rm j}$ of the location which is along the edge of the jet and closest to the line of sight (LOS) in Figure \ref{fig:beta}. One can see $\beta_{\rm app,j}$ reaches the maximum value $\beta_{\rm app,j,max}=\sqrt{\Gamma_{\rm j}^{2}-1}\approx\Gamma_{\rm j}$ when $\Gamma_{\rm j}=1/\sin(\theta_{v}-\theta_{j})\approx8.4$ \citep{Zhang2018}. The dimensionless apparent velocity of the flux centroid $\beta_{\rm app}$ trace $\beta_{\rm app,j}$ at early times, but $\beta_{\rm app}$ is clearly larger than $\beta_{\rm app,j}$ at late times. As shown in Figure \ref{fig:beta}, we also plot the apparent velocity of the FS flux centroid $\beta_{\rm app,FS}$ and the RS flux centroid $\beta_{\rm app,RS}$. There is almost no difference between $\beta_{\rm app,j}$, $\beta_{\rm app,FS}$, and $\beta_{\rm app,RS}$, which means that $\beta_{\rm app}$ is larger than $\beta_{\rm app,j}$, which is not because of the RS emission from the $e^{+}e^{-}$ pair wind. In order to illustrate this discrepancy, we show the predicted source radio images at 75 days, 207.4 days, and 230 days seen by the observer in Figure \ref{fig:image}. The black plus sign marks the location which is closest to the LOS, and the white plus sign marks the flux centroid. One can see the proper motion of the flux centroid is caused by three variations: the movement of the jet relative to observer, the variation of the radio afterglow image size, and the changes in the location of the flux centroid relative to the jet. All these variations would keep the flux centroid away from observer for an off-axis configuration. Therefore, a relatively large apparent velocity of the flux centroid is expected in late times.

\cite{Ghirlanda2019} reported GRB 170817A radio observations using VLBI with effective angular resolution is $1.5\times3.5$ mas, found that the source radio image appears compact and apparently unresolved, and estimated that the apparent size of the radio source is constrained to be smaller than 2.5 mas (90\% confidence level) at 207.4 days after the burst. In order to decide the size of our predicted source radio images, we adopt an elliptical Gaussian function to fit the images. The MCMC method is adopted to reach the best fit to the images. We derive the size of our predicted images from Result A and B at 207.4 days are about $2.4\times0.4$ mas and $3.4\times1.2$ mas, respectively. This is consistent with the expectation that a larger viewing angle leads to a smaller size of image due to the projection effect. The predicted size from Result A is consistent with the observational constraint, whereas the size from Result B is slightly larger. Although we cannot rule out Result B due to the size of the real observed image of our model is hard to decide, Result A is more promising than Result B.

\begin{deluxetable}{lcrcr}\label{tab:pars}
\tablecaption{Free parameters, priors, and best-fitting results in our model.}
\tablehead{
\colhead{Parameter}	&	\colhead{Prior A}	&	\colhead{Result A}	&	\colhead{Prior B}	&	\colhead{Result B}
}
\startdata
$\theta_j\rm\ (rad)$	&	$[0,1]$	&	$0.11_{-0.01}^{+0.01}$	&	$[0,1]$	&	$0.05_{-0.01}^{+0.01}$	\\
$\theta_v\rm\ (rad)$	&	$[0.23,0.7]$	&	$0.23_{-0.00}^{+0.00}$	&	$[0,0.7]$	&	$0.14_{-0.02}^{+0.02}$	\\
$\log[B_p\rm\ (G)]$	&	$[10,14]$	&	$13.20_{-0.03}^{+0.03}$	&	$[10,14]$	&	$13.25_{-0.04}^{+0.04}$	\\
$\log[E_{\rm K,iso}\rm\ (erg)]$	&	$[49,54]$	&	$52.27_{-0.89}^{+0.81}$	&	$[49,54]$	&	$52.87_{-0.79}^{+0.75}$	\\
$\log\Gamma_0$	&	$[1,3]$	&	$1.67_{-0.18}^{+0.12}$	&	$[1,3]$	&	$1.79_{-0.18}^{+0.13}$	\\
$\log[n_{1}\rm\ (cm^{-3})]$	&	$[-6,0]$	&	$-3.91_{-0.90}^{+0.81}$	&	$[-6,0]$	&	$-4.35_{-0.79}^{+0.79}$	\\
$\log\epsilon_{B,2}$	&	$[-7,-0.5]$	&	$-4.17_{-1.94}^{+2.04}$	&	$[-7,-0.5]$	&	$-3.93_{-1.75}^{+1.84}$	\\
$\log\epsilon_{B,3}$	&	$[-7,-0.5]$	&	$-5.51_{-0.08}^{+0.07}$	&	$[-7,-0.5]$	&	$-4.36_{-0.25}^{+0.33}$	\\
$\log\epsilon_{e,2}$	&	$[-7,-0.5]$	&	$-1.14_{-0.41}^{+0.41}$	&	$[-7,-0.5]$	&	$-2.47_{-0.55}^{+0.56}$	\\
$p_{2}$	&	$[2,3]$	&	$2.02_{-0.01}^{+0.01}$	&	$[2,3]$	&	$2.03_{-0.02}^{+0.03}$	\\
$p_{3}$	&	$[2,3]$	&	$2.21_{-0.03}^{+0.05}$	&	$[2,3]$	&	$2.16_{-0.02}^{+0.03}$	\\
\hline
$\chi^2/\rm{dof}$	&	-	&	$150.95/46$	&	-	&	$101.11/46$	\\
\enddata
\tablecomments{The uncertainties of the best-fitting parameters are measured as $1\sigma$ confidence ranges.}
\end{deluxetable}

Table \ref{tab:pars} shows the best-fitting parameters and their $1\sigma$ uncertainties by our fitting. The best-fitting parameters give $\theta_{j}\approx6^{\circ}$, which is consistent with the typical jet opening angle (e.g. \citealt{Frail2001,Wang2015,Wang2018}). The derived $\theta_{v}\approx13^{\circ}$ reaches the lower limit on the viewing angle by combining GW and EM constraints (e.g. \citealt{Finstad2018,Abbott2019}). An NS with $B_{p}\approx1.6\times10^{13}\rm\ G$ is needed in our fitting. Here $B_{p}$ is mainly determined by $\tau_{\rm sd}$, while $\tau_{\rm sd}$ is roughly equal to the peak time of the flux from the RS \citep{Geng2016}, and also roughly equal to the peak time of the afterglow ($\sim150$ days) in our fitting. Therefore, $B_{p}\sim10^{13}\rm\ G$ can be determined according to $\tau_{\rm sd}\approx150\rm\ days$. The isotropic kinetic energy $E_{\rm K,iso}\approx2\times10^{52}\rm\ erg$ corresponding to the true kinetic energy of jet $E_{\rm K}= E_{\rm K,iso}(1-\cos\theta_{j})/2\approx5.6\times10^{49}\rm\ erg$, which is comparable to the kinetic energy in the jet core inferred from structured jet models (e.g. \citealt{Hajela2019,Lamb2019,Troja2019,Ryan2020}). The initial Lorentz factor $\Gamma_{0}\approx47$ is lower than that obtained by the structured jet models. Besides, the microphysics parameters we derived are $\epsilon_{B,2}\approx7\times10^{-5}$, $\epsilon_{B,3}\approx3\times10^{-6}$, $\epsilon_{e,2}\approx0.07$, $\epsilon_{e,3}\approx1$, $p_{2}\approx2.0$, $p_{3}\approx2.2$. Figure \ref{fig:corner} displays the corner plots showing the results of our MCMC parameter estimation.
There is a strong degeneracy between $E_{\rm K,iso}$ and $n_{1}$, $E_{\rm K,iso}$ and $\epsilon_{B,2}$, $n_{1}$ and $\epsilon_{B,2}$, which leads to poor limitations on these parameters.

\section{Discussion \& Conclusions} \label{sec:discon}

Supposing the GRB 170817A prompt emission originates from internal dissipation of the jet energy (however, a cocoon shock breakout as the origin of the prompt emission was suggested by some other authors, e.g. \citealt{Kasliwal2017,Bromberg2018,Gottlieb2018}), the observed off-axis values of physical quantities of prompt emission can be corrected to an on-axis values via the angle-dependent Doppler factor. For the GRB duration $T_{90}$ and the peak of the GRB energy spectrum $E_{p}$, one has
\begin{equation}
\frac{T_{90,\rm off}}{T_{90,\rm on}}=\frac{E_{p,\rm on}}{E_{p,\rm off}}=\frac{\mathcal{D}(0)}{\mathcal{D}(\theta_{v}-\theta_{j})}=\frac{1-\beta\cos(\theta_{v}-\theta_{j})}{1-\beta},
\end{equation}
while for the isotropic $\gamma$-ray energy of prompt emission $E_{\gamma,\rm iso}$, one has
\begin{equation}
\frac{E_{\gamma,\rm iso,on}}{E_{\gamma,\rm iso,off}}\approx\left[\frac{\mathcal{D}(0)}{\mathcal{D}(\theta_{v}-\theta_{j})}\right]^2=\left[\frac{1-\beta\cos(\theta_{v}-\theta_{j})}{1-\beta}\right]^2.
\end{equation}
Using the observed quantities for GRB 170817A, $T_{90,\rm off}\approx2\rm\ s$, $E_{p,\rm off}\approx200\rm\ keV$, $E_{\gamma,\rm iso,off}=5.3\times10^{46}\rm\ erg$ \citep{Abbott2017b}, and our derived parameters, $\theta_{j}\approx0.11$, $\theta_{v}\approx0.23$, $\Gamma_{0}\approx47$, one can derive the on-axis values, $T_{90,\rm on}\approx0.06\rm\ s$, $E_{p,\rm on}\approx6.6\rm\ MeV$, $E_{\gamma,\rm iso,on}=5.7\times10^{49}\rm\ erg$.
The derived $T_{90,\rm on}$ falls into the $T_{90}$ distribution of sGRBs (e.g., \citealt{Kouveliotou1993}). The derived $E_{p,\rm on}$ is larger than typical value (several hundred keV), but is still consistent with the wide $E_{p}$ distribution of sGRBs (e.g. \citealt{Gruber2014}). After the correction, the energy released during the prompt emission phase of GRB 170817A is comparable to those sGRBs at low redshifts. For example, the short GRB 150101B jointly detected by \emph{Fermi}/GBM and \emph{Swift}/BAT, has a redshift $z=0.1343\pm0.0030$, which is the second nearby sGRB. At this redshift, the isotropic energy released in the $\sim10-1000$ keV energy band is $E_{\gamma,\rm iso}\approx1.3\times10^{49}\rm\ erg$ \citep{Fong2016}. GRB 160821B, the lowest redshift sGRB identified by \emph{Swift}, has a redshift $z=0.162$ \citep{Levan2016}, and its isotropic energy is $E_{\gamma,\rm iso}\approx2.1\times10^{50}\rm\ erg$ in the $8-1000$ keV range \citep{Lv2017}. The consistence between the on-axis values for GRB 170817A and typical values observed for sGRBs suggests that our model can reproduce the prompt emission of GRB 170817A.

The derived NS surface magnetic field strength is $B_{p}\sim10^{13}\rm\ G$ in our fitting, which is more than one order of magnitude higher than previous results. For example, \cite{Ai2018} constrained $B_{p}$ to be lower than $\sim10^{12}\rm\ G$\footnote{This constraint requires the ellipticity of the NS is smaller than $\sim10^{-4}$, which is consistent with the implicit condition of our model. By default, the rotational energy loss is dominated by magnetic dipole radiation rather than GW radiation in our model, i.e. the luminosity of GW emission should be less than the luminosity of magnetic dipole emission, which allows us to derive the upper limit of ellipticity $\lesssim5\times10^{-5}$ for $B_p\approx1.5\times10^{13}\rm\ G$.} if a long-lived NS survives after the merger. The constraint on $B_{p}$ is mainly contributed by the kilonova observations. One can directly derive the constraint on $B_{p}$ based on the ``Arnett Law'': the peak luminosity of the kilonova equals the heating rate at the peak time, i.e. $L_{\rm peak}\approx \dot{Q}(t_{\rm peak})$ \citep{Arnett1982}. The kilonova associated with GW170817 has a peak luminosity $L_{\rm peak}\sim10^{42}\rm\ erg$ at 0.5 days after merger. For a long-lived NS as post-merger remnant, the heating of the kilonova ejecta usually comes from two components: radioactive $r$-process heating and magnetic dipole spin-down heating, i.e. $\dot{Q}=\dot{Q}_{\rm ra}+\dot{Q}_{\rm md}$. Therefore, one can write $\dot{Q}_{\rm md}(t_{\rm peak})=\eta L_{\rm sd}(t_{\rm peak})\lesssim L_{\rm peak}$, where $L_{\rm sd}$ is the spin-down luminosity, and $\eta$ is a fraction of $L_{\rm sd}$ that is used to heat the ejecta. Since $\tau_{\rm sd}\gg t_{\rm\ peak}$, one has $L_{\rm sd,0}=9.64\times10^{44}B_{p,13}^{2}\lesssim\eta^{-1}L_{\rm peak}$. If one adopts $0.1<\eta<1$ as suggested by \cite{Yu2013}, then $B_{p}\lesssim10^{11}-10^{12}\rm\ G$, which is generally consistent with the result from \cite{Ai2018}. However, there are no simulations providing the exact value of $\eta$ so far, and $\eta$ could be much smaller than unity for the following reasons: (1) if a fraction of spin-down energy is reflected by the ejecta walls and the large pair optical depth through the nebula behind the ejecta, a low efficiency of the spin-down luminosity used to thermalization is expected, as suggested by \cite{Metzger2014}; (2) a fraction of spin-down energy could be converted to the kinetic energy of the kilonva ejecta rather than heat the ejecta (e.g. \citealt{Wang2016}); (3) The spin-down powered outflow could be collimated along the GRB jet direction due to the interaction between the outflow and the ejecta (e.g. \citealt{Bucciantini2012}). According to the derived $B_{p}$ and the peak luminosity of kilonova associated with GW170817, $\eta$ can be as low as $\sim10^{-3}$ in our estimation.

In this paper, we fit the multi-wavelength afterglow lightcurves and the VLBI proper motion of GRB 170817A based on the relativistic $e^{+}e^{-}$ pair wind model. The MCMC method is adopted to obtain the best-fitting parameters. We find that the overall quality of the fitting is good, indicating that our relativistic $e^{+}e^{-}$ pair wind model can explain the GRB 170817A afterglow. We obtain a set of best-fitting parameters using the prior of viewing angle ranging from 0.23 rad to 0.7 rad provided by the GW and EM obervations. The best-fitting value $\theta_v\approx0.23$ is close to the lower limit of the prior, indicating that our model prefers a smaller viewing angle. If one allows the prior of $0\leq\theta_{v}\leq0.7$, the best-fitting value of viewing angle can be as low as $\approx0.14$ rad. Besides, our model can also reproduce the prompt emission of GRB 170817A. A NS with $B_{p}\approx1.6\times10^{13}\rm\ G$ is needed in our fitting. Combining the derived $B_{p}$ and the kilonova observations, we find that the fraction of the spindown luminosity which is thermalized and available to power the kilonova can be as low as $\sim10^{-3}$. Our model can reproduce a late-time shallow decay in the X-ray lightcurve and predicts that the X-ray and radio flux will continue to decline in the next few years. In contrast, \cite{Hajela2021b} predicts that the X-ray and radio flux will keep increasing for at least a few years within the framework of the kilonova afterglow model, and the X-ray flux will remain constant or decay with index $-5/3$ in the fall-back accretion model \citep{Ishizaki2021}. Further observations of GRB 170817A afterglow may help verify or disprove our model.

\acknowledgments
We thank the referee for helpful comments that have allowed us to significantly improve our manuscript.
This work was supported by the National Key Research and Development Program of China (grant No. 2017YFA0402600), the National SKA Program of China (grant No. 2020SKA0120300), and the National Natural Science Foundation of China (grant No. 11833003).

\vspace{5mm}
\software{emcee \citep{emcee}}

\clearpage

\appendix

\section{Dynamics and radiation} \label{app:dyn}

There are four distinct regions when an ultra-relativistic $e^+e^-$ pair wind interact with the jet and medium: (1) the unshocked medium, (2) the shocked jet and medium, (3) the shocked wind, (4) the unshocked wind. In the following, the quantities of region $i$ are denoted by subscript $i$. The comoving- and observer-frame quantities are marked with and without a superscript prime ($^\prime$), respectively. Neglecting the internal structure of the blastwave \citep{Blandford1976}, and assuming that regions 2 and 3 move with the same Lorentz factor, i.e. $\Gamma_2=\Gamma_3=\Gamma$, the dynamics of such an FS-RS system can be solved by the energy conservation.

Let us first define a spherical coordinate system ($r,\theta,\phi$), where $r$ is the distance from the coordinate origin, $\theta$ and $\phi$ are the latitudinal and azimuthal angles, respectively. The GRB central engine is located at coordinate origin, and the GRB jet axis is along the direction of $\theta=0$. The observer lies on the $\phi=\pi/2$ plane, and $\theta_{v}$ is the angle between the LOS and the GRB jet axis. We divide the $\theta \in[0,\theta_{j}]$ and $\phi \in[0,2\pi]$ into $M$ and $N$ parts, so that the GRB jet can be discretized into $M\times N$ grids and each grid has a solid angle $d\Omega=\sin\theta d\theta d\phi$. For a top-hat jet model which is adopted in this paper, the dynamical evolution for each grid is identical.

For any grid, the dynamical evolution per unit solid angle can be described as follow. Considering radiative loss, the total kinetic energy of region 2 is
\begin{equation}\label{eq:E2}
E_{2}=(\Gamma-1)(M_{\rm ej}+m_{2}) c^{2}+(1-\epsilon_{2})\Gamma(\Gamma-1) m_{2} c^{2},
\end{equation}
where $M_{\rm ej}$ and $m_{2}$ are the rest masses of the initial GRB ejecta and FS swept-up medium per unit solid angle. $\epsilon_{2}=\epsilon_{e} t_{\rm syn}^{\prime-1}/(t_{\rm syn}^{\prime-1}+t_{\rm ex}^{\prime-1})$ is the radiation efficiency of region 2, where $\epsilon_{e}$ is the fraction of the shock internal energy that is partitioned to electrons, $t_{\rm syn}^{\prime}$ and $t_{\rm ex}^{\prime}$ are the synchrotron cooling timescale and the comoving-frame expansion timescale, respectively \citep{Dai1999}. Energy conservation requires that the change of the total kinetic energy of region 2 should be equal to the work done by region 3 to region 2 subtract a fraction of thermal energy which is radiated from region 2:
\begin{equation}\label{eq:dE2}
dE_{2}=R^{2} p_{3}^{\prime} dR-\epsilon_{2}\Gamma(\Gamma-1) dm_{2} c^{2},
\end{equation}
where the comoving pressure of region 3 should be calculated by
\begin{equation}\label{eq:p3}
p_{3}^{\prime}=(1-\epsilon_{3})(\Gamma_{34}-1)(\hat{\gamma}_{3}\Gamma_{34}+1)n_{4}^{\prime}m_{e}c^{2}.
\end{equation}
Here $\hat{\gamma}_{3}$ and $\epsilon_{3}$ are the adiabatic index and the radiation efficiency of region 3, $\Gamma_{34}=\Gamma_{43}\simeq(1/2)(\Gamma_{3}/\Gamma_{4}+\Gamma_{4}/\Gamma_{3})$ stands for the relative Lorentz factor between region 3 and 4, $n_{4}^{\prime}=n_{w}^\prime$ is the comoving $e^+e^-$ number density of region 4.
Combining equations (\ref{eq:E2}), (\ref{eq:dE2}), and (\ref{eq:p3}), the differential equation $d\Gamma/dR$ can be written in the form
\begin{equation}
\frac{d\Gamma}{dR}=\frac{R^{2}\left[(1-\epsilon_{3})(\Gamma_{34}-1)(\hat{\gamma}_{3}\Gamma_{34}+1)n_{4}^{\prime}m_{e}-(\Gamma^{2}-1) n_{1} m_{p}\right]}{M_{\rm ej}+\epsilon_{2} m_{2}+2(1-\epsilon_{2}) \Gamma m_{2}}.
\end{equation}
To solve the differential equation $d\Gamma/dR$, one needs to know the evolution of $m_{2}$ with distance $R$, which can be written as
\begin{equation}
\frac{dm_{2}}{dR}=R^{2} n_{1} m_{p}.
\end{equation}
Furthermore, the rest mass of region 3 is obtained by \citep{Dai2002}
\begin{equation}
\frac{dm_{3}}{dR}=R^{2}\left(\frac{\Gamma_{34} \beta_{34}}{\Gamma_{3} \beta_{3}}\right)n_{4}^{\prime} m_{\mathrm{e}}.
\end{equation}
Here $m_{3}$ is the rest mass of RS swept-up medium per unit solid angle, $\beta_{3}$ is the dimensionless speed of region 3, and $\beta_{34}\simeq(\Gamma_{3}^{2}-\Gamma_{4}^{2})/(\Gamma_{3}^{2}+\Gamma_{4}^{2})$ is the relative dimensionless speed between region 3 and 4.

In order to calculate the synchrotron and IC emission, we assume that a fraction of shock internal energy that is partitioned to magnetic fields $\epsilon_{B,i}$ and electrons $\epsilon_{e,i}$, and assume that electrons in regions 2 and 3 can be accelerated by FS and RS with a power-law distribution in energy $dN_{e,i}^{\prime}/d\gamma_{e,i}^{\prime}\propto\gamma_{e,i}^{\prime-p_{i}}(\gamma_{m,i}^{\prime}\leq\gamma_{e,i}^{\prime}\leq\gamma_{M,i}^{\prime})$. Here $p_{i}$ is the electron energy spectral index in region $i$, $\gamma_{m,2}^{\prime}=(m_{p}/m_{e})\epsilon_{e,2}(\Gamma_{2}-1)(p_{2}-2)/(p_{2}-1)$ and $\gamma_{m,3}^{\prime}=\epsilon_{e,3}(\Gamma_{43}-1)(p_{3}-2)/(p_{3}-1)$ are the comoving-frame minimum electron Lorentz factors in regions 2 and 3, $\gamma_{M,i}^{\prime}\simeq\sqrt{6\pi q_{e}/[\sigma_{T}B_{i}^{\prime}(1+Y_{i})}]\simeq10^{8}\sqrt{B_{i}^{\prime}(1+Y_{i})}$ is the maximum Lorentz factors in region $i$. The comoving magnetic field strength in regions 2 and 3 are $B_{2}^{\prime}=\sqrt{8\pi\epsilon_{B,2}n_{1}m_{\rm p}c^{2}(\Gamma_{2}-1)(\hat{\gamma_{2}}\Gamma_{2}+1)/(\hat{\gamma_{2}}-1)}$ and $B_{3}^{\prime}=\sqrt{8\pi\epsilon_{B,3}n_{4}^{\prime}m_{\rm e}c^{2}(\Gamma_{34}-1)(\hat{\gamma_{3}}\Gamma_{34}+1)/(\hat{\gamma_{3}}-1)}$, respectively. The Compton $Y$ parameter, defined by the ratio of the IC power to the synchrotron power, can be written as $Y_{i}=(-1+\sqrt{1+4\eta_{{\rm rad},i}\eta_{{\rm KN},i}\epsilon_{e,i}/\epsilon_{B,i}})/2$ \citep{Fan2006,He2009}, where $\eta_{{\rm rad},i}$ is the fraction of electron energy that is radiated, and $\eta_{{\rm KN},i}$ is the fraction of synchrotron photons below the KN limit frequency \citep{Nakar2007}.

The final electron energy has a broken power-law distribution, since the electrons are cooled by synchrotron and IC radiation. By comparing the electron minimum Lorentz factor $\gamma_{m}$ and the electron cooling Lorentz factor $\gamma_{c}$ that can be written as
\begin{equation}
\gamma_{{\rm c},i}^{\prime}=\frac{6\pi m_{e}c(1+z)}{\sigma_{T} B_{i}^{\prime2}\Gamma t_{\rm obs}(1+Y_{i})},
\end{equation}
two different regimes can be derived. If $\gamma_{c,i}^{\prime}\leq\gamma_{m,i}^{\prime}$, all the electrons have cooled, this is the fast cooling regime, one has
\begin{equation}
\frac{dN_{e,i}^{\prime}}{d \gamma_{e,i}^{\prime}} \propto\left\{\begin{array}{ll}
\gamma_{e,i}^{\prime -2} & \gamma_{c,i}^{\prime}\leq\gamma_{e,i}^{\prime}\leq\gamma_{m,i}^{\prime} \\
\gamma_{e,i}^{\prime -(p+1)} & \gamma_{m,i}^{\prime}<\gamma_{e,i}^{\prime}\leq\gamma_{M,i}^{\prime}
\end{array}\right..
\end{equation}
If $\gamma_{m,i}^{\prime}<\gamma_{c,i}^{\prime}$, only a fraction of electrons have cooled, this is the slow cooling regime, one has
\begin{equation}
\frac{dN_{e,i}^{\prime}}{d \gamma_{e,i}^{\prime}} \propto\left\{\begin{array}{ll}
\gamma_{e,i}^{\prime -p} & \gamma_{m,i}^{\prime}\leq\gamma_{e,i}^{\prime}\leq\gamma_{c,i}^{\prime} \\
\gamma_{e,i}^{\prime -(p+1)} & \gamma_{c,i}^{\prime}<\gamma_{e,i}^{\prime}\leq\gamma_{M,i}^{\prime}
\end{array}\right..
\end{equation}

Once the electron distribution is determined, the synchrotron radiation power per unit solid angle of each gird in the region $i$ at the frequency $\nu^{\prime}$ can be calculated as \citep{Rybicki1979}
\begin{equation}\label{eq:syn}
P_{i, \rm syn}^{\prime}(\nu^{\prime})=\frac{\sqrt{3} q_{e}^{3} B_{i}^{\prime}}{m_{e} c^{2}} \int_{\min(\gamma_{m, i}^{\prime},\gamma_{c, i}^{\prime})}^{\gamma_{M,i}^{\prime}} \frac{d N_{e, i}^{\prime}}{d \gamma_{e, i}^{\prime}} F\left(\frac{\nu^{\prime}}{\nu_{\rm ch}^{\prime}}\right) d \gamma_{e, i}^{\prime}.
\end{equation}
Here $\nu^{\prime}=(1+z) \nu_{\rm obs}/\mathcal{D}$ is the synchrotron frequency in the comoving frame, where $\nu_{\rm obs}$ is the observed frequency, $z$ is the redshift, and $\mathcal{D}\equiv 1/\Gamma(1-\beta\cos\alpha)$ is the Doppler factor with $\cos\alpha=\cos\theta\cos\theta_v+\sin\theta\sin\phi\sin\theta_v$ in our assumed geometrical setting ($\alpha$ is the angle between the velocity direction of each gird and the LOS). $q_{e}$ is the electron charge, $\nu_{\rm ch}^{\prime}=3 \gamma_{e, i}^{\prime 2} q_{e} B_{i}^{\prime} /(4 \pi m_{e} c)$ is the critical frequency of synchrotron radiation, $F\left(\nu^{\prime} / \nu_{\rm c}^{\prime}\right)=(\nu^{\prime} / \nu_{\rm ch}^{\prime}) \int_{\nu^{\prime} / \nu_{\rm ch}^{\prime}}^{+\infty} K_{5/3}(x) dx$, and $K_{5/3}(x)$ is the modified Bessel function of 5/3 order.

Besides the synchrotron radiation, we also consider the IC radiation from shock-accelerated electrons. The IC radiation include two parts: the self-Compton (SSC) radiation, and combined IC (CIC) radiation, i.e., photons in region $i$ are scattered by electrons in regions $j$ ($i\neq j$). The IC radiation power (per unit solid angle of each gird at the frequency $\nu^{\prime}$) of electrons in the region $i$ scatter photons from region $j$ can be calculated as ($i=j$ for SSC and $i\neq j$ for CIC; \citealt{Blumenthal1970,Yu2007})
\begin{equation}\label{eq:ic}
P_{i, \rm IC}^{\prime}(\nu^{\prime})= 3 \sigma_{\rm T} \int_{\gamma_{\min , i}^{\prime}}^{\gamma_{M, i}^{\prime}} \frac{dN_{e, i}^{\prime}}{d\gamma_{e, i}^{\prime}} d\gamma_{e, i}^{\prime} \int_{\nu_{s, j, \min}^{\prime}}^{\infty} d\nu_{s, j}^{\prime} \frac{\nu^{\prime} f_{\nu_{s, j}}^{\prime}}{4 \gamma_{e, i}^{\prime 2} \nu_{s, j}^{\prime 2}} g(x, y),
\end{equation}
where $\gamma_{\min,i}^{\prime}=\max[\min[\gamma_{c,i}^{\prime},\gamma_{m,i}^{\prime}],h\nu^{\prime}/(m_{e}c^{2})]$, $\nu_{s, j, \min}^{\prime}=\nu^{\prime}m_{e}c^{2}/[4\gamma_{e, i}^{\prime}(\gamma_{e,i}^{\prime}m_{e}c^{2}-h\nu^{\prime})]$. $g(x,y)=2y\ln y+(1+2y)(1-y)+\frac{x^{2}y^{2}}{2(1+xy)}(1-y)$ with $x=4\gamma_{e,i}^{\prime}h\nu_{s,j}^{\prime}/m_{e}c^{2}$ and $y=h\nu^{\prime}/[x(\gamma_{e, i}^{\prime} m_{e} c^{2}-h \nu^{\prime})]$.

Finally, we integrate equations (\ref{eq:syn}) and (\ref{eq:ic}) over the equal arrival time surface (EATS) to get the total flux density at the observed frequency $\nu_{\rm obs}$:
\begin{eqnarray}
F_{\nu_{\rm obs}}&=&\frac{1+z}{4 \pi D_{L}^{2}} \mathop{\int}_{\rm(EATS)} \mathcal{D}^{3} [P_{i, \rm syn}^{\prime}(\nu^{\prime})+P_{i, \rm IC}^{\prime}(\nu^{\prime})] d\Omega \\
&=&\frac{1+z}{4 \pi D_{L}^{2}} \mathop{\int_{0}^{\theta_j}\int_{0}^{2\pi}}_{\rm(EATS)} \mathcal{D}^{3} [P_{i, \rm syn}^{\prime}(\nu^{\prime})+P_{i, \rm IC}^{\prime}(\nu^{\prime})] \sin\theta d\theta d\phi.
\end{eqnarray}
For a given observed time $t_{\rm obs}$, the emission radius $R_{\alpha}$ at each grid (i.e. the EATS) is obtained by
\begin{equation}
t_{\rm obs}=(1+z)\int_{0}^{R_{\alpha}} \frac{1-\beta\cos\alpha}{\beta c} dr \equiv \rm const.
\end{equation}

\section{Proper motion of the flux centroid} \label{app:prop}

\begin{figure}
\centering
\includegraphics[width=0.3\textwidth,angle=0]{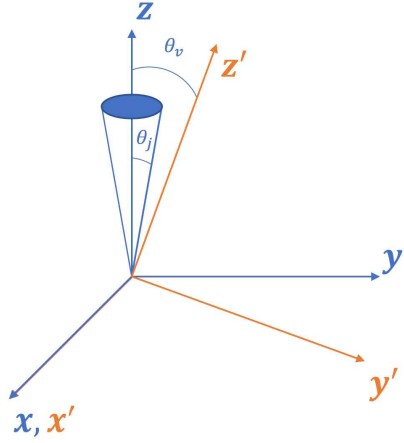}
\caption{Two coordinate systems before (blue) and after (orange) transformation. The jet direction along the $z$-axis, while the LOS is aligned with $z^{\prime}$-axis.
\label{fig:cord}}
\end{figure}

In order to calculate the proper motion of the flux centroid, one needs to project the position of each grid at each moment onto the plane perpendicular to the LOS. We first convert the spherical coordinate system $(R,\theta,\phi)$ to Cartesian coordinate system $(x,y,z)=(R\sin\theta\cos\phi,R\sin\theta\sin\phi,R\cos\theta)$. In the Cartesian coordinate system, the GRB jet axis is aligned with $z$ direction, and the observer lies on the $y-z$ plane. We perform a coordinate rotation about the $x$ axis from $(x,y,z)$ to $(x^{\prime},y^{\prime},z^{\prime})$, after which LOS is aligned with $z^{\prime}$ axis, and the angle between the $z$ axis and the $z^{\prime}$ axis is $\theta_{v}$, as shown in Figure \ref{fig:cord}. The coordinate transformation from $(x,y,z)$ to $(x^{\prime},y^{\prime},z^{\prime})$ is given by
\begin{eqnarray}
  x^{\prime}&=&x, \\
  y^{\prime}&=&y\cos\theta_{v}-z\sin\theta_{v}, \\
  z^{\prime}&=&z\cos\theta_{v}+y\sin\theta_{v}.
\end{eqnarray}
For a given observer time $t_{\rm obs}$, the projected position of each grid on the EATS is given by $\boldsymbol{r}_{g}^{\prime}=(x_{g}^{\prime},y_{g}^{\prime})=(R_{\alpha}\sin\theta\cos\phi,R_{\alpha}\cos\theta\sin\theta_{v}-R_{\alpha}\sin\theta\sin\phi\cos\theta_{v})$. Since the flux centroid is defined as the mean location of a distribution of flux density in space, the location of flux centroid on the plane perpendicular to the LOS can be expressed as
\begin{equation}
  \boldsymbol{r}_{\rm fc}^{\prime}=(x_{\rm fc}^{\prime},y_{\rm fc}^{\prime})=\frac{\int \boldsymbol{r}_{g}^{\prime} dF_{\nu_{\rm obs}}(x_{g}^{\prime},y_{g}^{\prime})}{\int dF_{\nu_{\rm obs}}(x_{g}^{\prime},y_{g}^{\prime})}=\frac{\int \boldsymbol{r}_{g}^{\prime} dF_{\nu_{\rm obs}}(x_{g}^{\prime},y_{g}^{\prime})}{F_{\nu_{\rm obs}}},
\end{equation}
where $dF_{\nu_{\rm obs}}(x_{g}^{\prime},y_{g}^{\prime})$ is the flux density of each grid on the EATS (one can easily calculate the surface brightness of each location via $dF_{\nu_{\rm obs}}(x_{g}^{\prime},y_{g}^{\prime})/dS_{\bot}$, where $dS_{\bot}$ is the area of each grid projected into the plane of the sky). Because of the jet is symmetric about the $y-z$ plane, the flux centroid can also described by $(0,y_{\rm fc}^{\prime})=(0, \int y_{g}^{\prime} dF_{\nu_{\rm obs}}/F_{\nu_{\rm obs}})$. For the observer, the angular position of each grid is $(0, y_{\rm fc}^{\prime}/D_A)$, where $D_A=D_L/(1+z)^2$ is the angular diameter distance. Finally, the average apparent velocity of the flux centroid over a time interval $t_{{\rm obs},j}-t_{{\rm obs},i}$ can be written as
\begin{equation}
v_{\rm app} = \frac{y_{\rm fc,j}^{\prime}-y_{\rm fc,i}^{\prime}}{t_{{\rm obs},j}-t_{{\rm obs},i}},
\end{equation}
and the dimensionless velocity $\beta_{\rm app}=v_{\rm app}/c$.

\section{Fitting results from Prior B} \label{app:PB}

\begin{figure}
\centering
\includegraphics[width=0.5\textwidth,angle=0]{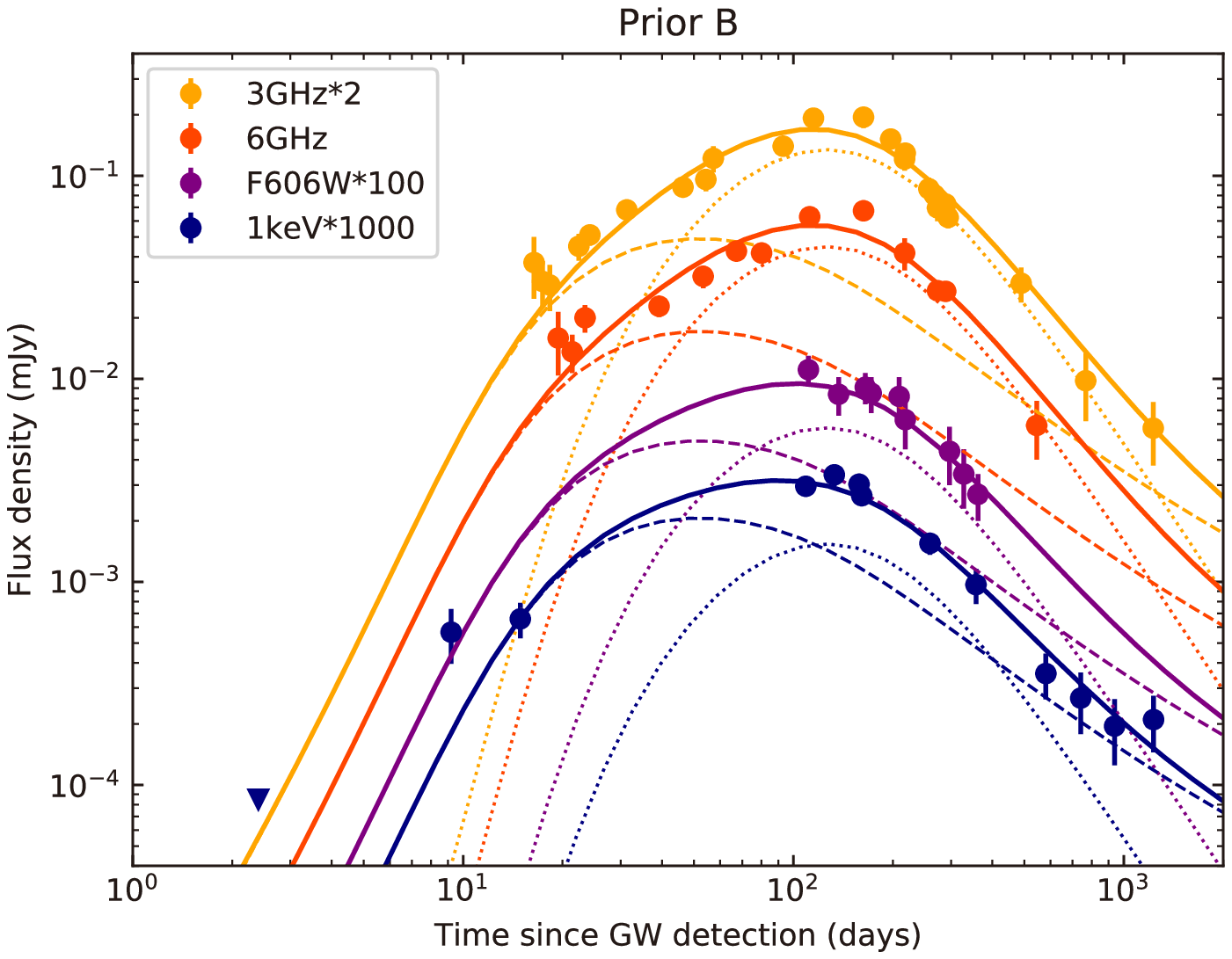}
\includegraphics[width=0.5\textwidth,angle=0]{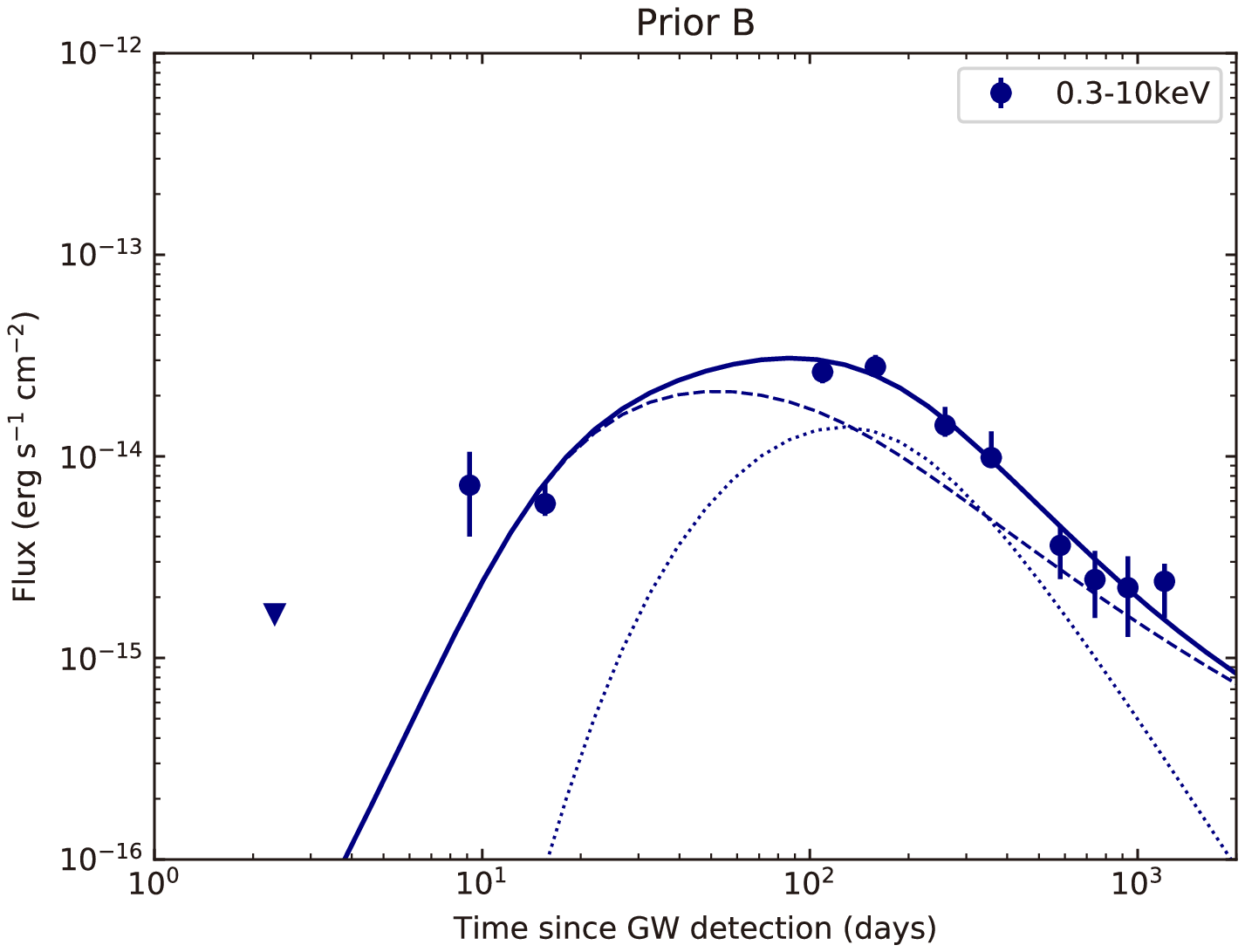}
\caption{Relativistic $e^+e^-$ pair wind model fit to the multi-wavelength afterglow lightcurves of GRB 170817A. The dashed and dotted curves represent the emission from the FS and RS, respectively. The solid curves are the total emission lightcurves.
\label{fig:multi2}}
\end{figure}

\begin{figure}
\centering
\includegraphics[width=0.5\textwidth,angle=0]{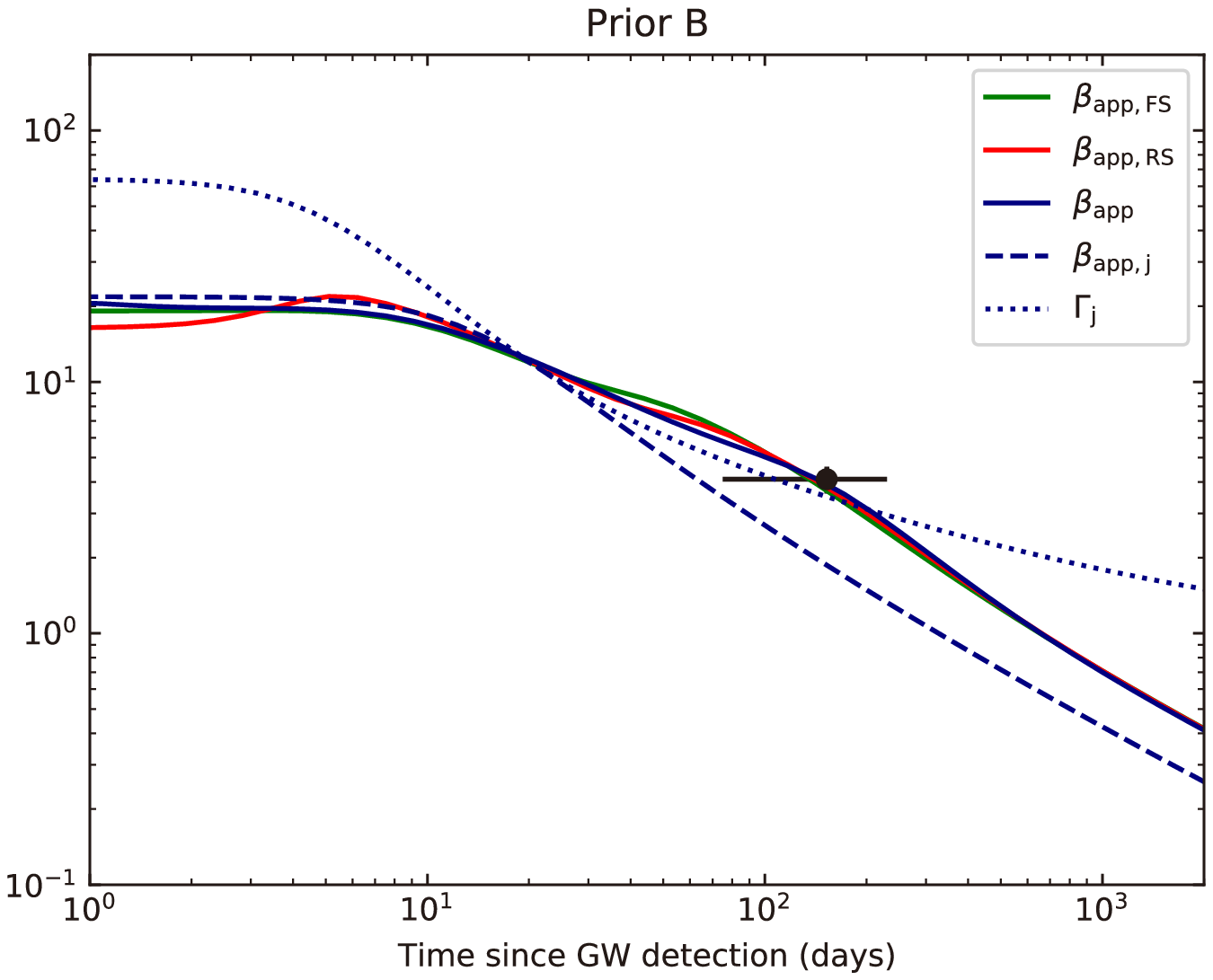}
\caption{Relativistic $e^+e^-$ pair wind model fit to the VLBI proper motion of GRB 170817A afterglow. The mean dimensionless apparent velocity of the radio afterglow source $\beta_{\rm app}=4.1\pm0.5$ between 75 days and 230 days after the merger from \cite{Mooley2018c} yields the data point used to fit. The green, red, and blue solid curve shows the evolution of dimensionless apparent velocity of the flux centroid based on our best-fitting parameters when considering FS emission, RS emission and total emission, respectively. The blue dashed and dotted curves denote $\beta_{\rm app,j}$ and $\Gamma_{\rm j}$.
\label{fig:beta2}}
\end{figure}

\begin{figure}
\centering
\includegraphics[width=0.5\textwidth,angle=0]{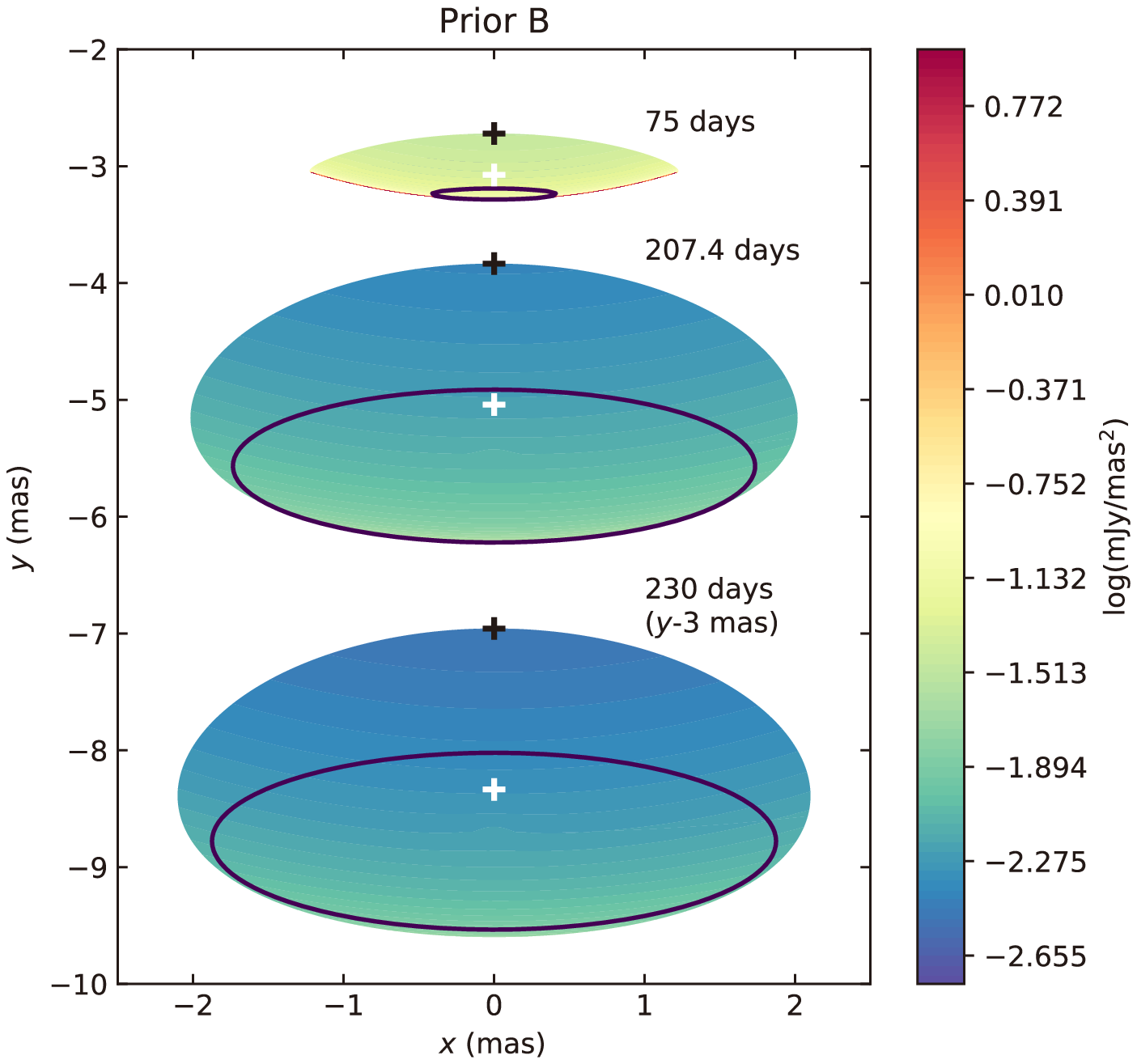}
\caption{The relative position of 4.5 GHz radio images at 75 days (upper), 207.4 days (middle), and 230 days (lower) seen by the observer. The white plus sign is the location of the flux centroid. The black plus sign is the location closest to the LOS. Purple ellipses are the best-fitting elliptical Gaussian showing the sizes (full width at half maximum) of the images.
\label{fig:image2}}
\end{figure}

\begin{figure*}
\centering
\includegraphics[width=1.0\textwidth,angle=0]{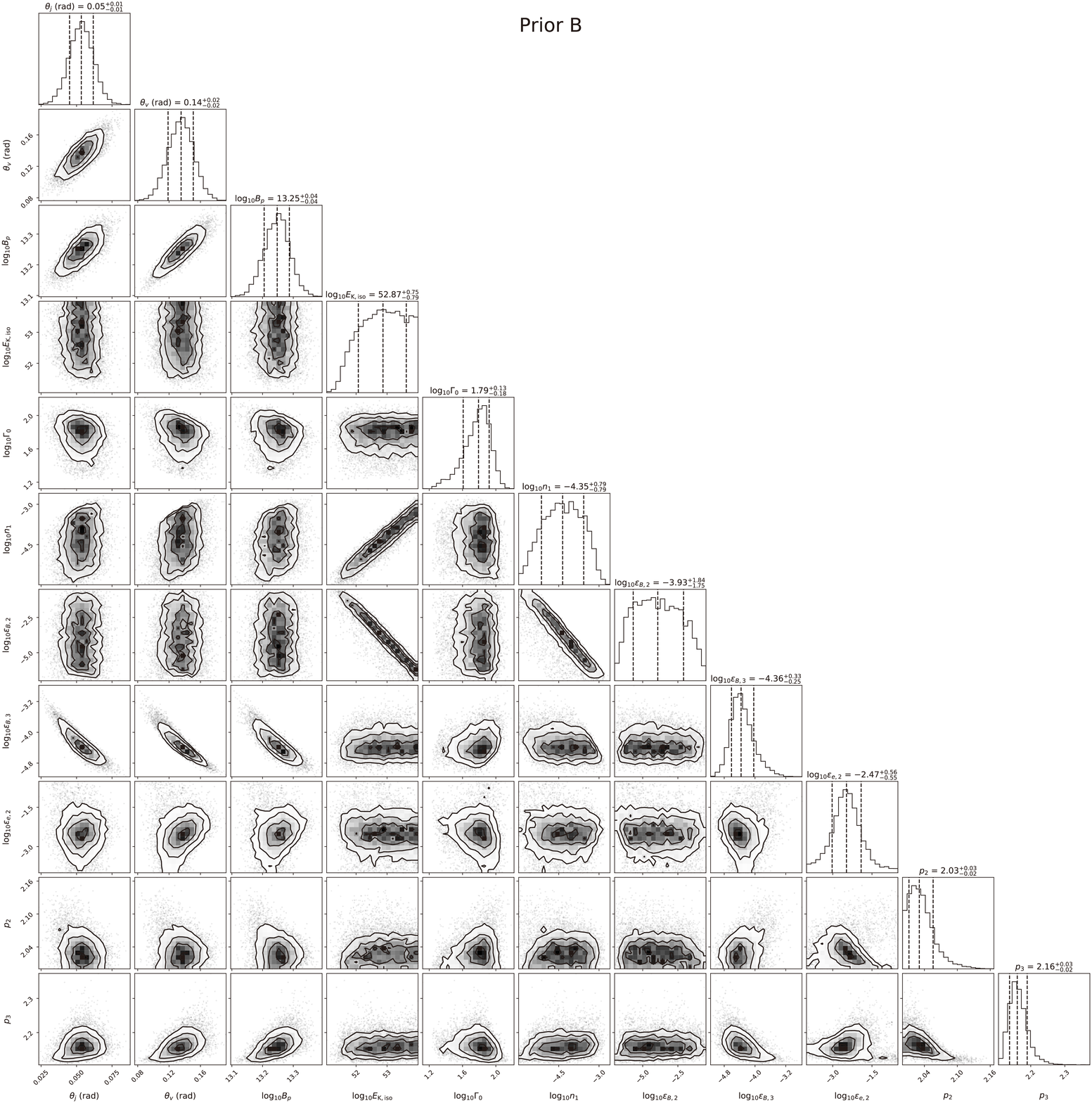}
\caption{A corner plot showing the results of our MCMC parameter estimation for the relativistic $e^+e^-$ pair wind model. Our best-fitting parameters and corresponding $1\sigma$ uncertainties are shown with the black dashed lines in the histograms on the diagonal.
\label{fig:corner2}}
\end{figure*}

We also show the fitting results from priors with $\theta_v\in[0,0.7]$ (Prior B) in Figure \ref{fig:multi2}, \ref{fig:beta2}, \ref{fig:image2}, \ref{fig:corner2}.

\clearpage


\bibliographystyle{aasjournal}
\end{document}